\documentclass[12pt]{iopart}

\newcommand \beq{\begin{eqnarray}}
\newcommand \eeq{\end{eqnarray}}
\usepackage{graphicx}
\usepackage{iopams} 
\usepackage{type1cm} 
\begin{document}

\title[Many Fermi polarons at nonzero temperature]{Many Fermi polarons at nonzero temperature}

\author{Hiroyuki Tajima$^1$ and Shun Uchino$^2$}

\address{$^1$Quantum Hadron Labolatory, RIKEN Nishina Center,
  Wako, Saitama 351-0198, Japan}
\address{$^2$Waseda Institute for Advanced Study, Waseda University, Shinjuku, Tokyo, 169-8050, Japan}

\ead{hiroyuki.tajima@riken.jp}
\vspace{10pt}
\begin{indented}
\item[]May 2018
\end{indented}

\begin{abstract}
An extremely polarized mixture of an ultracold Fermi gas is expected to  reduce to a Fermi polaron system,
which consists of a single impurity immersed in the Fermi sea of majority atoms.
By developing a many-body $T$-matrix theory, we investigate spectral properties of the polarized mixture
in experimentally relevant regimes in which  the system of finite impurity concentration
at nonzero temperature is concerned. 
We explicitly demonstrate presence of polaron physics in the polarized limit and discuss effects of
many polarons in an intermediate regime in a self-consistent manner.
By analyzing the spectral function at finite impurity concentration,
we extract the attractive and repulsive polaron energies.
We find that a renormalization of majority atoms via an interaction
with minority atoms and a thermal depletion of the impurity chemical potential are of significance to depict the many-polaron regime.
\end{abstract}

%\maketitle

\section{Introduction}

Understanding effects of impurities immersed in an environment is one of the key issues in physics.
In nuclear physics,  heavy hadrons in nuclear matter such as charm hadrons are now discussed in
context of impurity problems~\cite{HOSAKA}. In condensed matter physics, a number of impurities problems have been examined for a long time,
depending on conditions of impurities such as mobile or immobile
and presence or absence of a spin-exchange interaction~\cite{ashcroft}.
A particularly fundamental class of the problems is the polaron in which a mobile impurity interacts
with an environment~\cite{Landau,Feynman}. The concept of the polaron appears in a variety of the materials
such as metal, semiconductor, and superconductor systems~\cite{Devreese,0034-4885-72-6-066501}.

Currently, there is a growing interest in an ultracold atomic gas as a quantum simulator of polaron
physics~\cite{Schirotzek0902,Nascimbene2009,PhysRevA.85.023623,Kohstall1112,Eur.Phys.D.65.83,koschorreck2012,
PhysRevLett.117.055301,PhysRevLett.117.055302,Cetina96,PhysRevLett.118.083602,massignan}.  
The Feshbach resonance available in an ultracold atomic gas allows us to control an interaction
between impurity and bath and to investigate the strong coupling regime, which is generally challenging
in quantum many-body physics~\cite{RevModPhys.80.885}.
In addition, by using radio-frequency (rf) spectroscopy, we can address spectral properties of the systems
including excited branches~\cite{Le}. For example, rf spectroscopy experiments confirmed
existence of a repulsive polaron, which is a quasiparticle associated with a repulsive interaction and
is a metastable excited many-body state~\cite{Kohstall1112,Eur.Phys.D.65.83,koschorreck2012,PhysRevLett.117.055301,
PhysRevLett.117.055302,Cetina96,PhysRevLett.118.083602}.
The repulsive polaron also receives attention in terms of the realization of repulsive many-body states such as itinerant
ferromagnetism~\cite{Jo1521,PhysRevLett.106.050402,PhysRevLett.108.240404,PRL110.230401,LENSFM}.

Interpretations of polaron experiments in ultracold atomic gases are grounded on the theoretical analyses in which
the system with a single impurity at the zero temperature is assumed.
In the case of the Fermi polaron whose bath consists of fermions, due to such assumptions,
theoretical treatments such as variational methods~\cite{PhysRevA.74.063628,
Combescot2007,Punk2009,PhysRevA.80.033607,PhysRevA.81.041602,
PhysRevLett.106.166404,Trefzger2012}, 
$T$-matrix approximation~\cite{Combescot2008,Bruum2010,PhysRevA.85.021602,PhysRevA.85.033631}, functional renormalization~\cite{Schmidt:2011zu,PhysRevA.95.013612}, and diagrammatic Monte Carlo~\cite{Prokofev20081,Prokofev20082,Vlietinck2013,Kroiss2015,PhysRevA.94.051605} are successfully applied.
In the case of finite polarization, the polaron-polaron interaction is discussed \cite{PhysRevLett.97.200403,PhysRevLett.100.030401,PRA.79.013629,PRA.85.013605,2017arXiv170803410H}.
In reality, however, none of these theoretical assumptions are exactly satisfied
in corresponding experiments; the temperature is about from centesimal to few tenths of
the Fermi temperature \cite{Onofrio}
and impurity concentration is of the order of 10 percent.
Thus, it is important to directly analyze such regimes in terms of many-body calculations
accessible to 
the strong coupling regime.

In this paper, we examine spectral properties in the polarized mixture of an ultracold Fermi gas
with a many-body $T$-matrix theory, which allows us directly to plug in
the finite temperature and the impurity concentration effects.
We demonstrate that by shifting impurity concentration, the spectral function of impurities shows
crossover behaviors from a single polaron to many polarons. By analyzing the spectral function in detail, we
extract the polaron energy
as a function of impurity concentration.
We point out that a renormalization of majority atoms
due to minority atoms plays a crucial role in understanding the system at a finite density,
which has been overlooked in previous studies.
In addition, we show that the impurity chemical potential is largely affected by finite temperature effects compared to other quantities.
We also predict a quasiparticle-like peak in a high-energy regime
of the spectral function of majority atoms,
which cannot be captured with single-impurity theories and may be measured with rf spectroscopy.

%\textit{Formulation.}
\section{Formulation}
We consider the grand canonical Hamiltonian for the
two-component Fermi mixture interacting through the broad Feshbach resonance~\cite{RevModPhys.80.885}
(we set $\hbar=k_B=1$),
\beq
\label{eq1}
H=\sum_{\mathbf{k},\sigma}\xi_{\mathbf{p},\sigma}c^{\dagger}_{\mathbf{p},\sigma}
c_{\mathbf{p},\sigma}+g\sum_{\mathbf{p,q,k}}c^{\dagger}_{\mathbf{p},\uparrow}
c^{\dagger}_{\mathbf{q},\downarrow}c_{\mathbf{q+k},\downarrow}
c_{\mathbf{p-k},\uparrow},
\eeq
where $c_{\mathbf{p},\sigma}$ represents the fermionic annihilation operator with momentum $\mathbf{p}$
and pseudospin $\sigma=\uparrow,\downarrow$.
$\xi_{\mathbf{p},\sigma}=\frac{\mathbf{p}^2}{2m}-\mu_{\sigma}$ is the kinetic energy of atoms with mass $m$ measured from the chemical potential $\mu_{\sigma}$. The interatomic interaction is local and the coupling constant
 $g$ $(<0)$ can be characterized with the $s$-wave scattering length $a_s$~\cite{RevModPhys.80.885}. Notice that the system volume is taken to be unity.
Below, without loss of generality, we assume that $\uparrow (\downarrow)$ is
the majority (minority) spin.

We wish to examine the spectral function directly related to rf spectroscopy experiments, which is defined as
\beq
\label{eq2}
A_{\sigma}(\mathbf{p},\omega)=-\frac{1}{\pi}{\rm Im}G_{\sigma}(\mathbf{p},i\omega_n\rightarrow \omega+i\delta),
\eeq
where  the one-particle thermal Green's function is given by
\beq
\label{eq3}
G_{\sigma}(\mathbf{p},i\omega_n)=
\frac{1}{i\omega_n-\xi_{\mathbf{p},\sigma}-\Sigma_{\sigma}(\mathbf{p},i\omega_n)},
\eeq
with the self-energy $\Sigma_{\sigma}(\mathbf{p},i\omega_n)$. Here $\omega_n=(2n+1)\pi T$ is the fermionic Matsubara frequency ($T$ is the temperature) and $\delta$ is an infinitesimally small number. 
We note that the analytic continuation in Eq.~(\ref{eq2}) is numerically done by the Pad\'{e}  approximation with 
$\delta=10^{-3}\varepsilon_F$ where $\varepsilon_F$ is the Fermi energy of majority atoms (see also Appendix A).
From the definitions above, it follows that
the problem reduces to obtaining the self-energy that contains bare essentials of
the strongly interacting Fermi mixture.

To obtain the polaron energy $\omega_{\rm qp} \in \mathbb{R}$, we determine the pole $\omega_{\rm pole} \in \mathbb{C}$ of $G_{\downarrow}(\mathbf{p},\omega+i\delta)$ by solving a self-consistent equation%~\cite{note2}
\beq
\label{eq4}
\omega_{\rm pole}=\Sigma_{\downarrow}(\mathbf{p}=0,\omega_{\rm pole}+i\delta) - \mu_{\downarrow}.
\eeq
In general, $\omega_{\rm pole}$ locates on the complex plane of $\omega$, and especially in the case of repulsive polaron
near the unitarity in which the $s$-wave scattering length diverges,
the imaginary part of the self-energy is non-negligible. 
Therefore, we rewrite Eq. (\ref{eq4}) as
\beq
\label{eq4-1}
\omega_{\rm pole}+\mu_{\downarrow}=\omega_{\rm qp}-i\Gamma,
\eeq
with the decay rate $\Gamma \in \mathbb{R}$.
Here, $\omega_{\rm qp}$ and $\Gamma$ are related to the self-energy as
\beq
\label{eq4-2}
\omega_{\rm qp}={\rm Re}\Sigma_{\downarrow}(\mathbf{p}=0,\omega_{\rm qp}-\mu_{\downarrow}-i\Gamma+i\delta),
\eeq
\beq
\label{eq4-3}
\Gamma=-{\rm Im}\Sigma_{\downarrow}(\mathbf{p}=0,\omega_{\rm qp}-\mu_{\downarrow}-i\Gamma+i\delta).
\eeq
By solving the above two equations, we can obtain $\omega_{\rm qp}$ and $\Gamma$, respectively.

In addition, the chemical potential $\mu_{\sigma}$ is obtained from the so-called number equation 
\beq
\label{eq5}
n_{\sigma}(\mu_{\sigma})=T\sum_{\mathbf{p},i\omega_n}G_{\sigma}(\mathbf{p},i\omega_n),
\eeq
where $n_{\sigma}$ represents the particle density of atoms with the state
$\sigma$.
In this work, we define the impurity concentration $y$ as $y=n_{\downarrow}/(n_{\uparrow}+n_{\downarrow})$.
%%%%%%%%%%%%%%%%%%%%%%%%%%%%%%%%%%%%%%%%%%%%%%%%%%%%%%%%%%%%%%%%%%%%%
\begin{figure}[t]
\vspace{-0cm}
\begin{center}
\includegraphics[width=8cm]{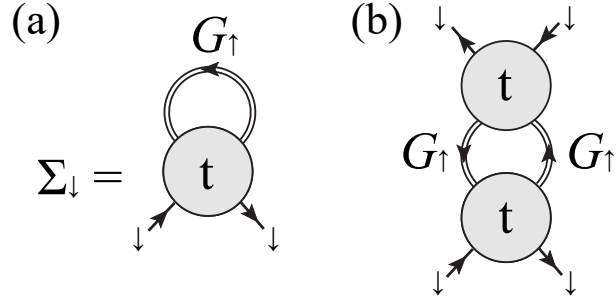}
\end{center}
\vspace{0cm}
\caption{(a) Diagrammatic expression for the ETMA self-energy of impurities $\Sigma_{\downarrow}$, where the non-selfconsistent $T$-matrix approximation is recovered if the dressed Green's function of medium $G_{\uparrow}$ (double solid line) is replaced by the non-interacting one. This self-energy includes the induced polaron-polaron interaction diagrammatically described by the process (b). Shaded circle represents the many-body $T$-matrix $t$.}
\label{fig1}
\end{figure}
%%%%%%%%%%%%%%%%%%%%%%%%%%%%%%%%%%%%%%%%%%%%%%%%%%%%%%%%%%%%%%%%%%%%%
\par
To obtain a reasonable self-energy,
we use many-body $T$-matrix theories, which are known to reproduce
fundamental properties
in spin-balanced~\cite{PhysRevB.63.224509,PhysRevB.66.024510,PhysRevA.80.033613,PhysRevA.80.063612}
and polaron limits~\cite{Combescot2008,Bruum2010,PhysRevA.85.021602,PhysRevA.85.033631}.
The simplest type of the $T$-matrix theories is the non-selfconsistent approximation whose
self-energy is composed of the bare Green's function.
However, such an approximation does not contain an interaction between impurities,
which is inevitable to discuss the finite impurity concentration case.
To overcome the drawback of the non-selfconsistent approximation, we adopt an extended $T$-matrix approximation
(ETMA)~\cite{PhysRevA.86.043622,PhysRevA.88.053621,PhysRevA.89.033617,PhysRevA.93.013610,
  PhysRevA.96.033614}, which contains the interaction between
impurities (see Fig.~\ref{fig1}) and therefore meets the purpose of the paper.
In this formalism, as diagrammatically shown in Fig. \ref{fig1}(a), the self-energy $\Sigma_{\sigma}(\mathbf{p},i\omega_n)$ is given by
\beq
\label{eqadd1}
\Sigma_{\sigma}(\mathbf{p},i\omega_n)=T\sum_{\mathbf{q},i\nu_n}t(\mathbf{q},i\nu_n)G_{-\sigma}(\mathbf{q}-\mathbf{p},i\nu_n-i\omega_n),
\eeq
where
\beq
\label{eqadd2}
t(\mathbf{q},i\nu_n)=\frac{g}{1+g\chi(\mathbf{q},i\nu_n)},
\eeq
is the many-body $T$-matrix ($\nu_n=2n\pi T$ is the bosonic Matsubara frequency). 
In Eq. (\ref{eqadd2}), the lowest-order-pair-correlation function $\chi(\mathbf{q},i\nu_n)$ is given by
\beq
\label{eqadd3}
\chi(\mathbf{q},i\nu_n)
&=& T\sum_{\mathbf{p},i\omega_j}G_{\uparrow}^{0}(\mathbf{p}+\mathbf{q},i\omega_j+i\nu_n)G_{\downarrow}^{0}(-\mathbf{p},-i\omega_n)\cr
&=&\sum_{\mathbf{p}}
\frac{1-f(\xi_{\mathbf{p+q},\uparrow})-f(\xi_{-\mathbf{p},\downarrow})}{\xi_{\mathbf{p+q},\uparrow}+\xi_{-\mathbf{p},\downarrow}-i\nu_n}
\eeq 
where $f(x)=1/(e^{x/T}+1)$ is the Fermi distribution function.
In Eq. (\ref{eqadd3}), $G_{\sigma}^0(\mathbf{p},i\omega_j)=1/(i\omega_j-\xi_{\mathbf{p},\sigma})$ is the bare Green's function.
Physically, $t(\mathbf{q},i\nu_n)$ describes superfluid fluctuations in the particle-particle channel \cite{PhysRevA.80.033613}.
Since the dressed Green's function $G_{\uparrow}$ in Eq. (\ref{eqadd1}) [or Fig. \ref{fig1}(a)] involves the self-energy $\Sigma_{\uparrow}$,
the polaron-polaron interaction process described by Fig. \ref{fig1} (b) is automatically included in the self-energy of minority atoms $\Sigma_{\downarrow}$.
We note that $\Sigma_{\sigma}(\mathbf{p},i\omega_n)$ is numerically obtained by self-consistently solving Eq. (\ref{eqadd1}) with calculating $\mu_{\sigma}$ from Eq. (\ref{eq5}), as shown in Fig.~\ref{figse}.
%\beq
%\label{eqadd3}
%\chi(\mathbf{q},i\nu_n)=T\sum_{\mathbf{p},i\omega_n}G^0_{\uparrow}(\mathbf{p}+\mathbf{q},i\omega_n+i\nu_n)G^0_{\downarrow}(-\mathbf{p},-i\omega_n),
%\eeq 
%where $G^0_{\sigma}(\mathbf{p},i\omega_n)=(i\omega-\xi_{\mathbf{p},\sigma})$ is the one-particle Green's function of a non-interacting Fermi atom.
\par
Recently, it was shown that the ETMA well reproduces thermodynamic properties in
spin-balanced systems~\cite{PhysRevA.95.043625,PhysRevX.7.041004}.
In what follows, we demonstrate that the ETMA also provides reasonable results on spectral properties
in the polarized system such as the polarons.
%We fix the temperature at $T=0.03T_{\rm F}$,
%where $T_{\rm F}$ is the Fermi temperature of majority atoms~\cite{note}.%\footnote{
  %We have checked that our results at $T=0.03T_{\rm F}$ stay almost unchanged up to
  % $T\approx0.1T_{\rm F}$, which is the temperature in the experiment~\cite{PhysRevLett.118.083602}.}.
In this work, we focus on the relevant parameter regimes to the recent experiments.
After discussing the comparison between our results and the previous works of experiments as well as theories
at the low temperature and impurity density regime,
we clarify effects of finite temperature and impurity density.
%\textit{Results.}---
\section{Result}
%%%%%%%%%%%%%%%%%%%%%%%%%%%%%%%%%%%%%%%%%%%%%%%%%%%%%%%%%%%%%%%%%%%%%
\begin{figure}[t]
\begin{center}
\includegraphics[width=14cm]{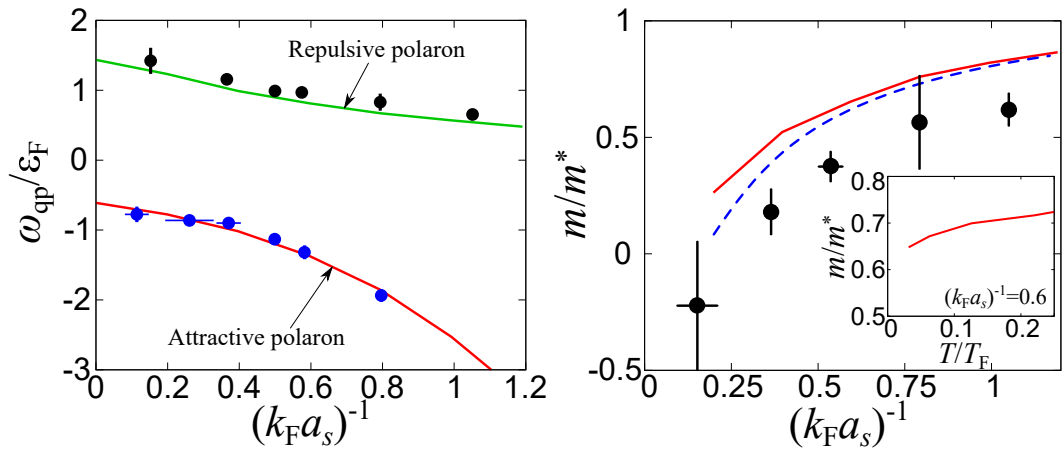}
\end{center}
\caption{(Left panel) Interaction dependence of the polaron energy near the zero impurity density limit ($y=n_{\downarrow}/(n_{\uparrow}+n_{\downarrow})\lesssim 10^{-3}$) at $T=0.03T_{\rm F}$.
  Solid lines show attractive (lower) and repulsive (upper) polaron energies calculated by the ETMA.
  The dots represent the experimental results in $^6$Li Fermi gases~\cite{PhysRevLett.118.083602}.
(Right panel) The effective mass of repulsive polarons $m^{*}$ near the zero impurity limit.
The solid line shows our result with the ETMA. The dashed line is the result in the previous work~\cite{Eur.Phys.D.65.83}.
The black dots are observed effective masses in Ref.~\cite{PhysRevLett.118.083602}.
The inset shows the calculated temperature dependence of $m^{*}$ at $(k_{\rm F}a_s)^{-1}=0.6$.}
\label{fig2}
\end{figure}
%%%%%%%%%%%%%%%%%%%%%%%%%%%%%%%%%%%%%%%%%%%%%%%%%%%%%%%%%%%%%%%%%%%%%
%At first, we verify reliability of the ETMA through a comparison with
%experimental measurements at zero impurity density limit.
We first show how our many-body $T$-matrix theory works well even
in the zero-impurity density limit at the low temperature through a comparison between
our numerical results and
the recent experimental measurements~\cite{PhysRevLett.118.083602} as well as previous theoretical studies.
In our formalism, the zero-impurity density limit is achieved by putting the large chemical potential difference $\mu_{\uparrow}-\mu_{\downarrow}$ such that the impurity concentration $y=n_{\uparrow}/(n_{\uparrow}+n_{\downarrow})\lesssim 10^{-3}$ is enough small.
The left panel of Fig.~\ref{fig2} shows the attractive or repulsive polaron energy $\omega_{\rm qp}$ as
a function of inverse scattering length $(k_{\rm F}a_s)^{-1}$ with the Fermi momentum of majority atoms
$k_{\rm F}$.
In our calculation, the temperature is fixed at $T=0.03T_{\rm F}$ (where $T_{\rm F}$ is the Fermi temperature of majority atoms).
Our results show good agreements with recent experimental results in $^6$Li Fermi gases~\cite{PhysRevLett.118.083602}. We note that while the experiment~\cite{PhysRevLett.118.083602}
has been done at a bit higher
impurity density and higher temperature compared with our theoretical input,
the differences do not lead to significant consequences on the polaron energy as discussed below.
In addition, in the zero impurity density limit, the ETMA reduces to the 
non-selfconsistent $T$-matrix approximation,
which is known to  describe polaron properties quantitatively,
since the majority one-particle Green's function $G_{\uparrow}(\mathbf{p},i\omega_n)$
in the ETMA
reduces to non-interacting one
$G_{\uparrow}^0(\mathbf{p},i\omega_n)=1/(i\omega_n-\xi_{\mathbf{p},\sigma})$~\cite{2017arXiv170803410H}
in the zero impurity density limit.
Thus, our approach based on the ETMA turns out to be a natural extension of the non-selfconsistent $T$-matrix approximation
with a single impurity to discuss finite temperature and density
in the Fermi polaron system.
\par
Our result of the effective mass $m^{*}$ subtracted from $G_{\downarrow}(\mathbf{p},\omega+i\delta)$ near the single impurity limit is
shown in the right panel of Fig.~\ref{fig2} and is consistent with the previous work~~\cite{Eur.Phys.D.65.83}. 
The small difference between the previous and our works comes from the finite temperature effects as shown in the inset of the right panel of Fig.~\ref{fig2}.
It is quite natural that $m^{*}$ decreases with increasing the temperature since the temperature effects gradually suppress the interaction effects.
This is the reason why our calculated $m^{*}$ at $T=0.03T_{\rm F}$ is larger than that of the previous work obtained at $T=0$.
On the other hand, the experimental results~\cite{PhysRevLett.118.083602} show heavier effective masses than our evaluation in spite of the fact that the experimental temperature $T=0.1T_{\rm F}$ is higher than our case.
We also numerically checked that the effect of a finite-impurity density does not lead such significant difference.
The large mass renormalization in the recent experiment~\cite{PhysRevLett.118.083602}
cannot be explained by finite temperature or impurity density effect by means of the ETMA.
\par
%%%%%%%%%%%%%%%%%%%%%%%%%%%%%%%%%%%%%%%%%%%%%%%%%%%%%%%%%%%%%%%%%%%%%
\begin{figure}[t]
\begin{center}
\includegraphics[width=14cm]{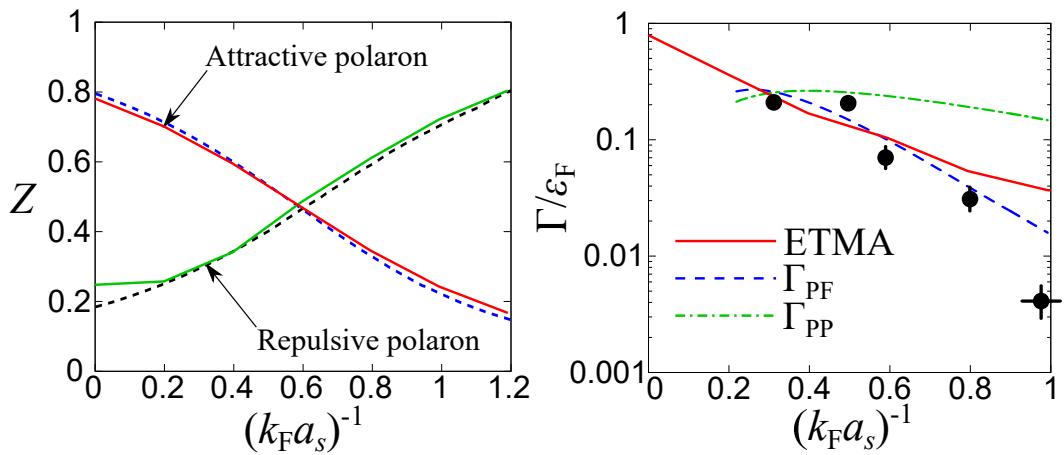}
\end{center}
\caption{The left panel shows residue $Z$ of each polaron calculated by the ETMA (solid line) in the zero-impurity limit at $T=0.03T_{\rm F}$ and the functional renormalization group (FRG) in Ref.~\cite{Schmidt:2011zu} (dashed line).
The right panel is the interaction dependence of the decay rate of repulsive polarons $\Gamma$ at $T=0.03T_{\rm F}$.
In this figure, the black dots are experimental results~\cite{PhysRevLett.118.083602}.
$\Gamma_{\rm PF}$ and $\Gamma_{\rm PP}$ are the decay rate at $T=0$ of polaron-to-bare-atom and polaron-to-polaron processes, respectively \cite{Eur.Phys.D.65.83}.
}
\label{figZG}
\end{figure}
%%%%%%%%%%%%%%%%%%%%%%%%%%%%%%%%%%%%%%%%%%%%%%%%%%%%%%%%%%%%%%%%%%%%%
The left panel of Fig.~\ref{figZG} shows the residue $Z$ of minority Green's function at $\omega=\omega_{\rm pole}$, which
is calculated as \cite{Schmidt:2011zu}
\beq
\label{eqZ}
Z^{-1}=-\left.\frac{\partial}{\partial \omega}G_{\downarrow}^{-1}(\mathbf{p}=0,\omega+i\delta)\right|_{\omega=\omega_{\rm pole}}.
\eeq
Our results of $Z$ for attractive and repulsive polarons show good agreement with the theoretical study based on the functional renormalization group at $T=0$ \cite{Schmidt:2011zu}, which non-perturbatively involves higher order corrections such as three-body process.
From this comparison, one can find that the reside $Z$ is essentially described well by the ladder-approximation scheme at the single-impurity limit. 
\par
However, the decay rate of repulsive polarons $\Gamma$ obtained from Eqs. (\ref{eq4-2}) and (\ref{eq4-3}) is generally smaller compared to FRG results \cite{Schmidt:2011zu} since our calculation does not incorporate the effect of three-body decay associated with atom-dimer scatterings \cite{Petrov} as well as the decay to attractive polarons, which can be considered by replacing $G_{\downarrow}^0$ in $\chi(\mathbf{q},i\nu_n)$ with dressed one $G_{\downarrow}$~\cite{Eur.Phys.D.65.83}. 
Since $G_{\downarrow}^{0}$ is concerned, the ETMA may reproduce the decay rate of polaron-to-bare-atom transition $\Gamma_{\rm PF}$ rather than that of polaron-to-polaron transition $\Gamma_{\rm PP}$ calculated in the previous work at the single-impurity limit with exactly
$T=0$~\cite{Eur.Phys.D.65.83}.
Although the correct physical process may be the latter, the former is closer to the experimental result.
In addition, our result involves finite temperature effects which enhance the decay of the quasi-particles~\cite{2017arXiv170803410H},
which is visible in the weak repulsive interacting regime where the collisional effects are relatively small.
%We note that the repulsive polaron spectrum near the unitarity limit shows a large broadened peak and the precise determination of its properties are difficult in our formalism. 
\par
%%%%%%%%%%%%%%%%%%%%%%%%%%%%%%%%%%%%%%%%%%%%%%%%%%%%%%%%%%%%%%%%%%%%%
\begin{figure}[t]
\begin{center}
\includegraphics[width=8cm]{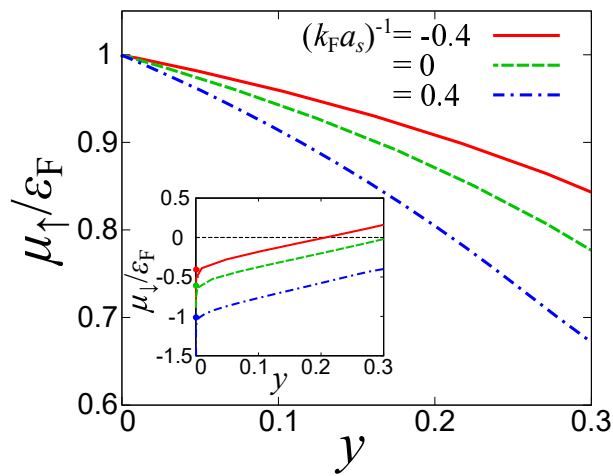}
\end{center}
\caption{Calculated chemical potential of majority atoms as a function of impurity concentration
  $y$ at $T=0.03T_{\rm F}$.
  Inset shows results of minority atoms. In each figure, we use the same line-style
  at each impurity concentration.The circle represents the attractive polaron energy $\omega_{\rm qp}^{\rm a}$ at the single-impurity limit.}
\label{fig3}
\end{figure}
%%%%%%%%%%%%%%%%%%%%%%%%%%%%%%%%%%%%%%%%%%%%%%%%%%%%%%%%%%%%%%%%%%%%%
We next look at how impurity concentration $y$
affects the chemical potential $\mu_{\sigma}$.
We note that $\mu_{\uparrow}=\varepsilon_{\rm F}$ in the single impurity case at
$T=0$.
However, as shown in Fig.~\ref{fig3},
$\mu_{\uparrow}$ deviates from the Fermi energy and
decreases with increasing $y$ due to the self-energy shift
${\rm Re}\Sigma_{\uparrow}(\mathbf{p},\omega+i\delta)$ associated
with the strong pairing interaction.
This renormalization effect on majority atoms becomes more remarkable
when the pairing interaction gets stronger. 
%In addition, the chemical potential of minority atoms $\mu_{\downarrow}$
%increases with $y$ in each interaction strength.
%We note that $\mu_{\downarrow}$ shows a discontinuity at $y=0$ since $\mu_{\downarrow}\rightarrow-\infty$ in the absence of impurities ($y=0$), while $\mu_{\downarrow}$ is finite for more than or equal to a
%single impurity ($y\rightarrow 0$) \cite{Punk2009}.
Furthermore, the shifts of $\mu_{\sigma}$ are not explained by the simple mean-field shift $\Sigma_{\sigma}^{\rm MF}=\frac{4\pi a_s}{m}n_{-\sigma}$, since the scattering length $a_s$
diverges near the unitarity limit.
However, the shift of $\mu_{\uparrow}$ is proportional to $n_{\downarrow}$ even at $(k_{\rm F}a_s)^{-1}=0$ at small $y$.
By using the linear fitting with respect to $n_{\downarrow}/n_{\uparrow}$ in the small impurity-density regime ($y<0.2$), we obtain
\beq
\label{eqmu}
\mu_{\uparrow}=\varepsilon_{\rm F}\left[1-0.526\frac{n_{\downarrow}}{n_\uparrow}\right]\equiv\varepsilon_{\rm F}\left[1-0.526\frac{y}{1-y}\right].
\eeq 
Surprisingly, as pointed out in Ref.~\cite{Kinnunen}, this shift is the same-order of the mean-field shift with $a=1/k_{\rm F}$ given by
\beq
\label{eqmu2}
\Sigma_{\rm MF}(a=1/k_{\rm F})=\frac{4\pi}{mk_{\rm F}}n_{\downarrow}\simeq0.424\frac{n_{\downarrow}}{n_{\uparrow}}\varepsilon_{\rm F}.
\eeq
Since the chemical potential plays a crucial role in the thermodynamics of a unitary Fermi gas in which 
$\mu_{\uparrow}/\varepsilon_{\rm F}$ in the unpolarized case takes a universal constant called Bertsch parameter \cite{Bertsch}, we expect that
the origin of pre-factor $0.526$ in the second term of the right hand side of Eq. (\ref{eqmu}) would be important in terms of the thermodynamics of the many polarons.
We emphasize that these renormalization effects cannot be captured with
single-impurity theories.
The renormalization is of the order of a tenth of the Fermi energy
in the typical cold-atom experiments whose impurity concentration is $0.1$ to $0.3$.
We expect that such a significant shift 
can be measured with the state-of-the-art precision thermodynamic measurement~\cite{PhysRevX.7.041004}.
\par
The inset of Fig. \ref{fig3} shows the impurity chemical potential $\mu_{\downarrow}$,
which monotonically increases with increasing $y$ and decreases with increasing the interaction strength.
At the zero temperature, $\mu_{\downarrow}$ is equivalent to the attractive polaron energy $\omega_{\rm qp}^{\rm a}$ at $y\rightarrow 0$, 
since $\mu_{\downarrow}=E_{N_{\downarrow}=1} -E_{N_{\downarrow}=0}$ is defined as the energy needed to add an impurity with zero momentum to the system where $E_{N_{\downarrow}}$ ($N_{\downarrow} \in \mathbb{Z}$) is the energy in the presence of $N_{\downarrow}$ impurities.
Indeed, this definition is equivalent to $\mu_{\downarrow}=\left(\frac{\partial E}{\partial n_{\downarrow}}\right)_{S}$ at the thermodynamic limit, where $E$ and $S$ are the internal energy and entropy, respectively.
At a finite temperature, however, we have to carefully notice the difference between $\mu_{\downarrow}$ and $\omega_{\rm qp}^{\rm a}$.
An important point is that at a finite temperature there is the contribution from thermal excited states with nonzero
momenta in addition to one from the ground state with the zero momentum.
%%%%%%%%%%%%%%%%%%%%%%%%%%%%%%%%%%%%%%%%%%%%%%%%%%%%%%%%%%%%%%%%%%%%%
\begin{figure}[t]
\begin{center}
\includegraphics[width=8cm]{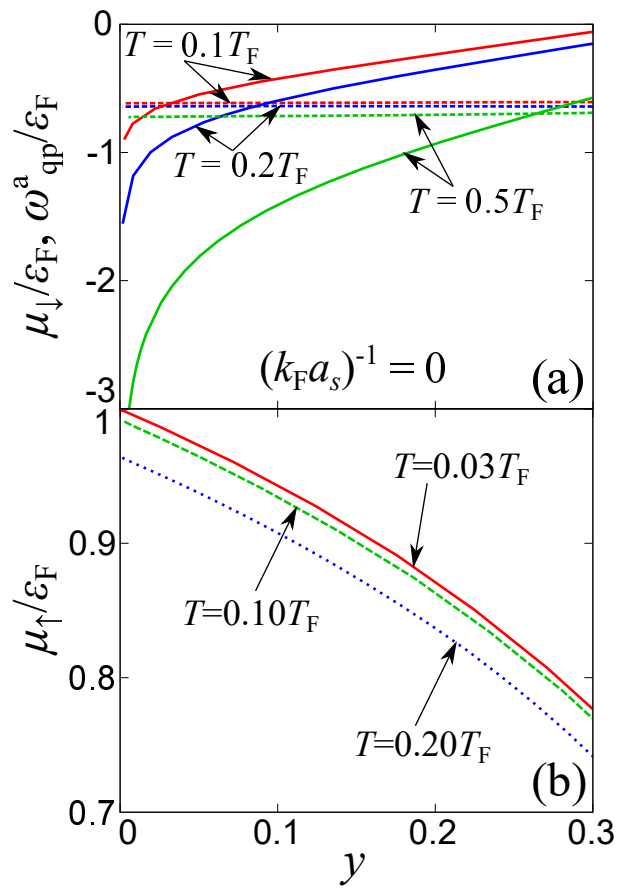}
\end{center}
\caption{Impurity concentration dependence of (a) $\mu_{\downarrow}$ (solid line) and $\omega_{\rm qp}^{\rm a}$ (dashed line) at $T=0.1T_{\rm F}$, $0.2T_{\rm F}$ and $0.5T_{\rm F}$, and (b) $\mu_{\uparrow}$ at $T=0.03T_{\rm F}$, $0.10T_{\rm F}$ and $0.20T_{\rm F}$.
In each figure, the interaction strength is set at $(k_{\rm F}a_s)^{-1}=0$.
}
\label{figmw}
\end{figure}
%%%%%%%%%%%%%%%%%%%%%%%%%%%%%%%%%%%%%%%%%%%%%%%%%%%%%%%%%%%%%%%%%%%%%
Figure \ref{figmw} (a) shows the impurity chemical potential $\mu_{\downarrow}$ and $\omega_{\rm qp}^{\rm a}$of the unitarity limit as a function of $y$ at several temperatures.
In general, $\mu_{\downarrow}$ is smaller than $\omega_{\rm qp}^{\rm a}$ in the small impurity density region ($y\simeq 0$).
In addition, $\mu_{\rm \downarrow}$ decreases with increasing the temperature, whereas $\omega_{\rm qp}^{\rm a}$ slightly shifts due to the temperature effects.
Except for the strong-coupling regime beyond polaron-molecule (or polaron-BEC) transition, the number equation of impurities for $\mu_{\downarrow}$ can approximately be given by
\beq
\label{eqt}
n_{\downarrow}\simeq\sum_{\mathbf{p}}Z_{\rm a}f\left(\frac{\mathbf{p}^2}{2m_{\rm a}^{*}}-\mu_{\downarrow}+\omega_{\rm qp}^{\rm a}\right),
\eeq
where $Z_{\rm a}$ and $m_{\rm a}^{*}$ are the residue and effective mass of an attractive polaron, respectively.
For simplicity, we neglect the decay rate of an attractive polaron as well as the repulsive branch.
At $T=0$, the solution of Eq. (\ref{eqt}) for the low impurity density limit ($n_{\downarrow}\rightarrow 0$) is apparently $\mu_{\downarrow}=\omega_{\rm qp}^{\rm a}$ since the Fermi distribution function $f(x)$ becomes a step function $\theta(-x)$.
On the other hand, at finite temperature, such solution have to be $\mu_{\downarrow}\rightarrow -\infty$ 
because the summation over momenta in Eq. (\ref{eqt}) involves the contribution from high momentum region associated with the finite temperature.
This large negative $\mu_{\downarrow}$ reflects the fact that a few polarons at finite temperature behave as a classical Boltzmann ensemble.
Indeed, if one measures the temperature by using the Fermi temperature of impurities $T_{\rm F,\downarrow}$, one can obtain
\beq
\label{eqt2}
\frac{T}{T_{\rm F,\downarrow}}=\left(\frac{n_{\uparrow}}{n_{\downarrow}}\right)^{2\over3}\frac{T}{T_{\rm F}},
\eeq 
which diverges in the limit of $n_{\downarrow}\rightarrow 0$ with fixed $T/T_{\rm F}$.
In contrast, the region where $\mu_{\downarrow}>\omega_{\rm qp}^{\rm a}$ at the large impurity density can be regarded as the Fermi degenerate regime of attractive polarons.
In this case, they make a soft Fermi surface with the effective Fermi energy $\varepsilon_{\rm F}^{\rm p}=\mu_{\downarrow}-\omega_{\rm qp}^{\rm a}$.
To access such a regime, the temperature must be much smaller than $T_{\rm F,\downarrow}=(n_{\downarrow}/n_{\uparrow})^{2\over3}T_{\rm F}$.
In Fig. \ref{figtf}, we summarize
the different regimes in the Fermi polaron system.
We also note that the curves shown in Fig. \ref{figtf}
are shifted below if the effective mass is considered, since $T_{\rm F,\downarrow}$ is generally in inverse proportion to the effective mass.
%%%%%%%%%%%%%%%%%%%%%%%%%%%%%%%%%%%%%%%%%%%%%%%%%%%%%%%%%%%%%%%%%%%%%
\begin{figure}[t]
\begin{center}
\includegraphics[width=8cm]{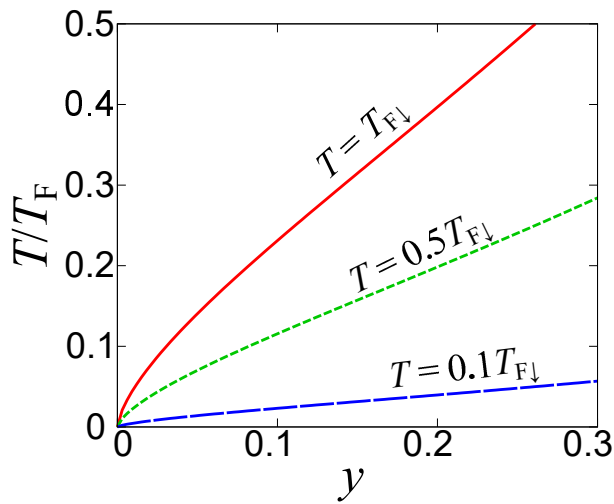}
\end{center}
\caption{Different regimes in the Fermi polaron system obtained from Eq. (\ref{eqt2}). The region above the red curve is approximated as
a classical Boltzmann gas. On the other hand, the region below the blue curve is described with a theory at $T=0$.
In between, there exists a soft Fermi surface in which the finite temperature effect is significant.
}
\label{figtf}
\end{figure}
%%%%%%%%%%%%%%%%%%%%%%%%%%%%%%%%%%%%%%%%%%%%%%%%%%%%%%%%%%%%%%%%%%%%%
\par
We note that in contrast to $\mu_{\downarrow}$,
the spectral property of the attractive polaron at the single-impurity limit is relatively robust against
the finite temperature effects, since it is related to the thermodynamic property of majority atoms.
At $n_{\downarrow}\rightarrow 0$ where $G_{\uparrow}(\mathbf{p},\omega+i\delta)\simeq\delta(\omega-\xi_{\mathbf{p},\uparrow})$, the self-energy of impurities after the analytic continuation is given by
\beq
\label{eqt3}
\Sigma_{\downarrow}(\mathbf{p},\omega+i\delta)=\sum_{\mathbf{q}}\int_{-\infty}^{\infty}d\nu
\frac{}{}A_{t}(\mathbf{q},\nu)
\frac{b(\nu)+f(\xi_{\mathbf{q}-\mathbf{p},\uparrow})}{\omega+i\delta+\xi_{\mathbf{q}-\mathbf{p},\uparrow}-\nu}
\eeq
where $A_{t}(\mathbf{q},\nu)=-\frac{1}{\pi}{\rm Im}t(\mathbf{q},i\nu_n\rightarrow\nu+i\delta)$ is the spectral function of a diatomic pair
and $b(x)=1/(e^{x/T}-1)$ is the Bose distribution function.
The finite temperature effects in Eq. (\ref{eqt3}) originate from mainly $f(\xi_{\mathbf{q}-\mathbf{p},\uparrow})$ and $\mu_{\uparrow}\simeq\varepsilon_{\rm F}\left[1-\frac{\pi^2}{12}\left(\frac{T}{T_{\rm F}}\right)^2\right]$~\cite{Feynman} [see Fig. \ref{figmw} (b)] far away from the BEC critical point of molecules.
In this way, one can find that spectral polaron properties such as $\omega_{\rm qp}$ determined by Eq. (\ref{eq4}) is deeply related to how majority fermions are affected by the temperature.
We also note that the large negative $\mu_{\downarrow}$ does not notably affect $\Sigma_{\downarrow}(\mathbf{p}=0,\omega+i\delta)$ since $\mu_{\downarrow}$ in Eq. (\ref{eqt3}) is included in only the molecular branch $A_{t}(\mathbf{q},\nu)$. 
%%%%%%%%%%%%%%%%%%%%%%%%%%%%%%%%%%%%%%%%%%%%%%%%%%%%%%%%%%%%%%%%%%%%%
\begin{figure}[t]
\begin{center}
\includegraphics[width=8cm]{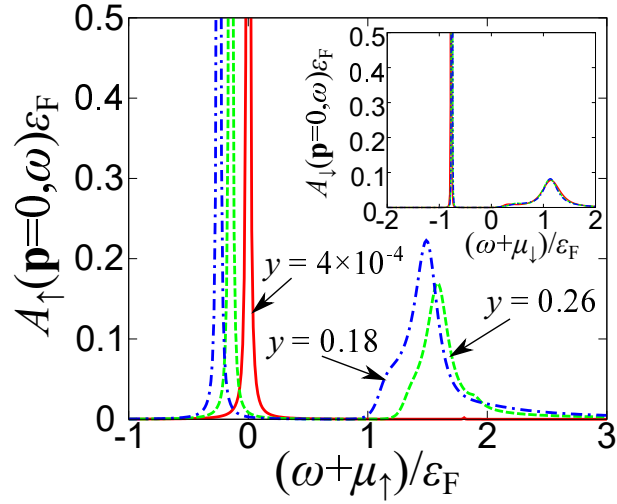}
\end{center}
\caption{Spectral function $A_{\uparrow}(\mathbf{p}=0,\omega)$ of majority atoms at
  $(k_{\rm F}a_s)^{-1}=0.2$.
  with different impurity concentrations. Inset shows that of minority atoms.
  The solid, dot-dashed, and dashed lines represent the result of $y=4\times10^{-4}$, $0.18$, and $0.26$, respectively.
  The temperature is fixed at $T=0.03T_{\rm F}$.
  In each figure, we use the same line style in each impurity concentration.
}
\label{fig4}
\end{figure}
%%%%%%%%%%%%%%%%%%%%%%%%%%%%%%%%%%%%%%%%%%%%%%%%%%%%%%%%%%%%%%%%%%%%%
\par
A renormalization of majority atoms is also visible in the spectral function
$A_{\uparrow}(\textbf{p}=0,{\omega})$.
In Fig.~\ref{fig4}, we show the spectral function at $y=4\times10^{-4}, 0.18$ and
$0.26$ at $(k_{\rm F}a_s)^{-1}=0.2$.
It turns out that the stable pole position shifts toward the lower energy with
increasing $y$ due to the shift of $\mu_{\uparrow}$.
From Eq. (\ref{eq3}), the shift of the peak in Fig. \ref{fig4} is directly related to the change of the self-energy of majority atoms as given by ${\rm Re}\Sigma_{\uparrow}(\mathbf{p}=0,\omega+i\delta)$. 
This is nothing but the renormalization effect of majority atoms.
In addition, we find that a metastable peak associated with the upper branch appears
at finite impurity concentration even in the spectral function of majority atoms.
The presence of such a peak originates from the upper peak of the minority Green's function
that is explicitly contained in the self-energy of majority atoms.
We also confirm that the metastable-peak structure is enhanced in the vicinity of
the strong coupling limit.
By considering that the intensity of such an upper peak in the majority spectral function is
comparable to that in minority spectral function,
its experimental validation with rf spectroscopy is promising. 
%%%%%%%%%%%%%%%%%%%%%%%%%%%%%%%%%%%%%%%%%%%%%%%%%%%%%%%%%%%%%%%%%%%%%
\begin{figure}[t]
\begin{center}
\includegraphics[width=8cm]{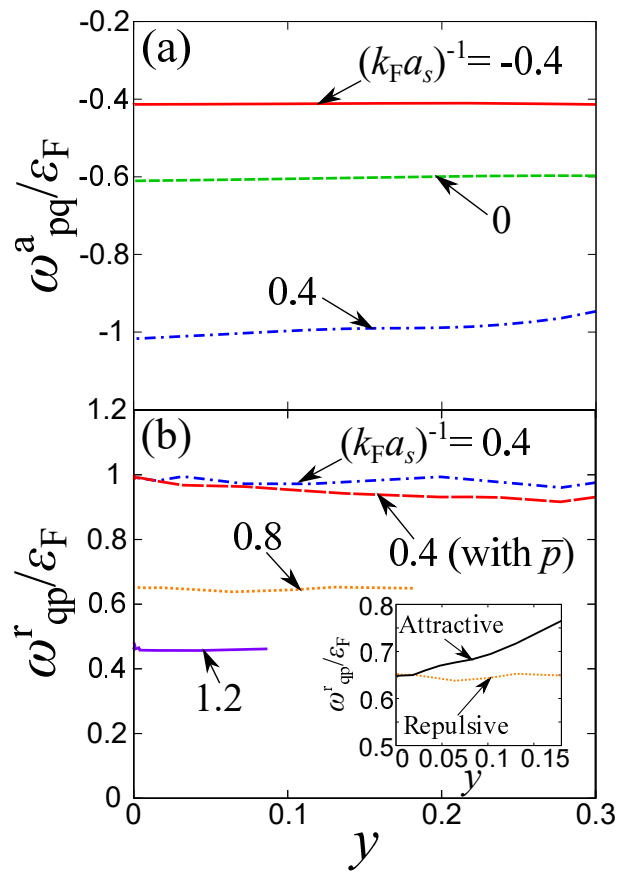}
\end{center}
\caption{Impurity concentration dependence of (a) attractive and (b)
  repulsive polaron energies at $T=0.03T_{\rm F}$.
  In the panel (b), the dashed line shows the repulsive polaron energy $\omega_{\rm qp}^{\rm r}(\bar{p})$ with the initial state momentum $\bar{p}$ at $(k_{\rm F}a_s)^{-1}=0.4$.
  The inset of (b) shows a comparison between the attractive (solid line) and repulsive (dotted line) polaron energies at $(k_{\rm F}a_s)^{-1}=0.8$, where we set an offset ($=2.5\varepsilon_{\rm F}$) on the attractive polaron energy.
}
\label{fig5}
\end{figure}
%%%%%%%%%%%%%%%%%%%%%%%%%%%%%%%%%%%%%%%%%%%%%%%%%%%%%%%%%%%%%%%%%%%%%
\par
On the other hand, in contrast to majority atoms, 
the shift of the spectral function $A_{\downarrow}(\mathbf{p}=0,\omega)$ of minority atoms by the finite density is small as shown in the inset of Fig.~\ref{fig4}.
In Fig.~\ref{fig5}~(a), we show impurity concentration dependence of the attractive polaron energy $\omega_{\rm qp}^{\rm a}$ obtained from Eq.~(\ref{eq4}) at several interaction strength. 
We find that $\omega_{\rm qp}^{\rm a}$ is almost independent of $y$
from the weak coupling region to unitary region.
However, in the strong coupling region [$(k_{\rm F}a_s)^{-1}=0.4$ in Fig.~\ref{fig5}(a)],
the polaron energy turns to slightly increase with increasing $y$.
We argue that this indicates the presence of the polaron-polaron interaction,
which is indeed known to be positive by means of the Fermi liquid
theory~\cite{PhysRevLett.97.200403,PhysRevLett.100.030401,2017arXiv170803410H}.
One can interpret that the polaron-polaron interaction effect is visible due to the increase of pairing interaction that  overcomes the finite temperature effect.
Indeed, the increase of $\omega_{\rm qp}^{\rm a}$ at $(k_{\rm F}a_s)^{-1}=0.4$ starts around $y\simeq 0.2$, where $T/T_{\rm F,\downarrow} \lesssim0.1$ estimated by Eq. (\ref{eqt2}) and the attractive polarons are in the deep quantum degenerate regime.
In this sense, the precise determination of $\mu_{\downarrow}$ is very important even from such viewpoint for the polaron-polaron interaction.
\par
In Fig.~\ref{fig5}~(b), we show the calculated repulsive polaron energy $\omega_{\rm qp}^{\rm r}$ as a function of $y$ in the strong coupling region [$(k_{\rm F}a_s)^{-1}=0.4, 0.8$ and $1.2$].
In addition, the inset of Fig. \ref{fig5} (b)
is the comparison between $y$-dependence of attractive and repulsive polaron energies
at $(k_{\rm F}a_s)^{-1}=0.8$, where we set an offset ($=2.5\varepsilon_{\rm F}$) on the attractive polaron energy.
These results indicate that  
the repulsive polaron energy does not represent any noteworthy
behavior related to the polaron-polaron interaction, which is consistent
with the recent experiment~\cite{PhysRevLett.118.083602}.
While solely from our numerical data it is difficult to pinpoint the reason of the difference from
the prediction of the Fermi liquid theory, the followings could be conceivable:
(i) smallness of the polaron-polaron interaction due to the Pauli blocking,
(ii) short lifetime of the repulsive polaron (typically of the order of the Fermi time),
(iii) finite temperature effect as is the case with attractive polaron.
%However, our estimation shows that the decay time of the repulsive polaron in the strong coupling regime
%s of the order of the Fermi time [$\simeq O(1)\varepsilon_{\rm F}^{-1}$] in contrast to the stable attractive polaron.
%By considering that it takes about 2.7 collisions for $\uparrow$ and $\downarrow$ particles
%to thermalize~\cite{PhysRevLett.70.414,PhysRevLett.101.100401} and the collision time
%in the strong coupling regime is again about the Fermi time,
\par
We note that we stop the calculations of $\omega_{\rm qp}^{\rm r}$ at the superfluid instability point,
which can be identified by the so-called Thouless criterion~\cite{THOULESS1960553}, 
\beq
\label{eq6}
[t(\mathbf{q}=0,i\nu_n=0)]^{-1}=0. 
\eeq
%Here, $t(\mathbf{q},i\nu_n)$ is the particle-particle many-body $T$-matrix with
%the momentum $\mathbf{q}$ and bosonic Matsubara frequency $\nu_n=2\pi nT$.
At the fixed temperature, the Thouless criterion is more likely to be satisfied in the regime
$(k_{\rm F}a_s)^{-1}\gtrsim0$, where the transition temperature of the superfluid is higher and increases
with increasing $y$.
To correctly describe the superfluid phase transition in a strongly interacting spin-imbalanced Fermi gas, we have to consider the existence of the first order phase transition and the phase separation \cite{Sheehy,Parish}.
In this paper, we avoid such a regime by focusing on lower impurity concentration.
We also note that although the realization of the Fulde-Ferrell-Larkin-Ovchinnikov (FFLO) state \cite{FF,LO} has been predicted in a uniform polarized Fermi gas \cite{Sheehy}, such an exotic superfluid state is known to be unstable against superfluid fluctuations \cite{Shimahara,Ohashi} (note however Ref. \cite{Zwerger}).   
\par
%%%%%%%%%%%%%%%%%%%%%%%%%%%%%%%%%%%%%%%%%%%%%%%%%%%%%%%%%%%%%%%%%%%%%
\begin{figure}[t]
\begin{center}
\includegraphics[width=8cm]{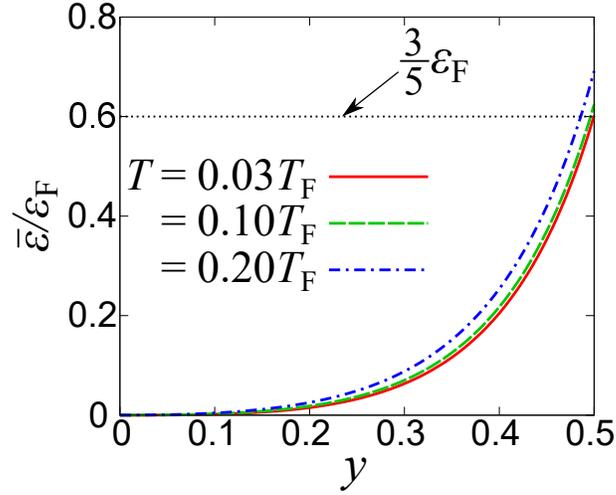}
\end{center}
\caption{Estimated thermal average of the impurity energy $\bar{\varepsilon}$ in the initial state.
The dotted line indicate the internal energy density of an ideal Fermi gas at $T=0$.}
\label{fige}
\end{figure}
%%%%%%%%%%%%%%%%%%%%%%%%%%%%%%%%%%%%%%%%%%%%%%%%%%%%%%%%%%%%%%%%%%%%%
Furthermore, to address the more detailed experimental situation,
we consider the effect of initial-state momentum of impurities in the rf spectrum measurement~ \cite{PhysRevLett.118.083602}.
We first estimate the averaged momentum of impurities $\bar{p}$ by assuming that the initial state is a non-interacting uniform Fermi gas.
The thermal average of the impurity energy $\bar{\varepsilon}$ is defined as
\beq
\label{eqa}
\bar{\varepsilon}=\frac{1}{n_{\downarrow}}\sum_{\mathbf{p}}\frac{\mathbf{p}^2}{2m}f\left(\frac{\mathbf{p}^2}{2m}-\mu_{\rm i}\right),
\eeq
where $\mu_{\rm i}$ is the chemical potential of the initial state impurities, obtained by the solving
\beq
\label{eqa2}
n_{\downarrow}=\sum_{\mathbf{p}}f\left(\frac{\mathbf{p}^2}{2m}-\mu_{\rm i}\right).
\eeq
From the above equations, we can obtain $\bar{p}=\sqrt{2m\bar{\varepsilon}}$.
Figure \ref{fige} shows the impurity concentration dependence of $\bar{\varepsilon}$.
In the relevant region of the experimental impurity density ($0.1\lesssim y \lesssim 0.3$) and temperature ($T\simeq 0.1T_{\rm F}$),
it is quite small compared to the trapped case reported in the Supplemental Material of Ref.~\cite{PhysRevLett.118.083602}.
In the presence of $\bar{p}$, the repulsive polaron energy is obtained from
\beq
\label{eqa3}
\omega_{\rm qp}^{\rm r}(\bar{p})-i\Gamma=\Sigma_{\downarrow}(\mathbf{\bar{p}},\omega_{\rm qp}^{\rm r}(\bar{p})-\mu_{\downarrow}-i\Gamma+i\delta).
\eeq
The dashed line in Fig. (\ref{fig5}) shows calculated $\omega_{\rm qp}^{\rm r}(\bar{p})$ at $(k_{\rm F}a_s)^{-1}=0.4$.
As expected, the fiinite $\bar{p}$ leads to the negative shift of $\omega_{\rm qp}^{\rm r}(\bar{p})$ compared to $\omega_{\rm qp}^{\rm r}(\bar{p}=0)$.
In the experimental paper, it is estimated that this negative shift is given by $-\left(1-\frac{m}{m^{*}}\right)\bar{\varepsilon}$ with $\bar{\varepsilon}=O(10^{-1}\varepsilon_{\rm F})$~\cite{PhysRevLett.118.083602}.
However, in our case, $\bar{\varepsilon}$ is smaller than $10^{-1}\varepsilon_{\rm F}$ in the relevant region, and the estimated shift is also smaller than $O(10^{-2}\varepsilon_{\rm F})$.
This result indicates the importance of effects of a harmonic trap potential to see the mass renormalization effects from the $y$-dependence of polaron energies.
Since the harmonic trap enhance the finite temperature effects due to the inhomogeneous density profile \cite{PhysRevA.96.033614},
it may also be related to the suppression of effects of polaron-polaron interaction in the experiment.
%%%%%%%%%%%%%%%%%%%%%%%%%%%%%%%%%%%%%%%%%%%%%%%%%%%%%%%%%%%%%%%%%%%%%
%\begin{figure}[t]
%\begin{center}
%\includegraphics[width=8cm]{fig6.eps}
%\end{center}
%\caption{ (a) The attractive polaron energy $\omega_{\rm qp}^{\rm a}+\mu_{\downarrow}$ and (b) the chemical potential of majority atoms $\mu_{\uparrow}$ as a function of $y$ at $T=0.03T_{\rm F}$, $0.10T_{\rm F}$, and $0.20T_{\rm F}$ at the unitarity limit ($1/k_{\rm F}a_s=0$).
%}
%\label{fig6}
%\end{figure}
%%%%%%%%%%%%%%%%%%%%%%%%%%%%%%%%%%%%%%%%%%%%%%%%%%%%%%%%%%%%%%%%%%%%%
%Figure \ref{fig6} shows the temperature dependence of the attractive polaron energy $\omega_{\rm qp}^{\rm a}+\mu_{\downarrow}$ (a) and the chemical potential of majority atoms $\mu_{\uparrow}$ (b) at the unitarity limit [$(k_{\rm F}a_s)^{-1}=0$].
%From this result, we find that the temperature dependence of polaron properties is not so strong in the very low temperature regime ($T\lesssim 0.1T_{\rm F}$).
%While $\mu_{\uparrow}$ shows a monotonic decrease due to the thermal depletion with increasing $T$, the attractive polaron energy at $T=0.20T_{\rm F}$ exhibits a slight suppression in the case of the large impurity concentration ($y\gtrsim 0.2$),
%which might be originated from a non-Fermi liquid behavior due to many-body effects with thermal fluctuations \cite{2017arXiv170803410H}.
 
\par  
%\textit{Conclusion.}---
\section{Conclusion}
We have theoretically investigated Fermi polarons at finite impurity concentration and finite temperature
within the framework of the many-body $T$-matrix theory, which can also
describe polaron properties in the zero impurity density and zero temperature limits.
Our results show quantitative or semi-quantitative agreement with current experiments as well as previous works based on single polaron theories at zero temperature.
\par
We have pointed out that majority atoms are affected by the strong pairing interaction with impurities.
In particular, we have showed the renormalization effects on the chemical potential as well as quasi-particle spectral function of majority atoms.
In the case of minority atoms, the finite temperature effects play a crucial role in their thermodynamic properties such as chemical potential.
It is also related to the quantum degeneracy of attractive polarons, which leads to the competition between finite temperature effects and the polaron-polaron interaction.
The renormalization of the majority chemical potential and the thermal depletion of minority chemical potential can be observed by recent precise thermodynamic measurements.
In addition, we have predicted 
the appearance of the metastable peak in the high-energy region of majority spectral function.
A detailed study on such a metastable many-body state is an interesting future work.
Also, metastable peak structure in the spectral function of majority atoms can be detected by rf spectrum measurements.
\par
We have also extracted the polaron energy as a function of impurity concentration to discuss the
polaron-polaron interaction.
We have found that in the strong coupling region at a low temperature, although the polaron-polaron interaction is visible in the lower branch, this effect is much weaker in the upper branch.
In addition, we also have clarified that the mass-renormalization effect on the polaron energy in the uniform case is smaller compared to the case of trapped gas clouds, by considering the initial-state momentum of impurities.
\par
In this paper, we have emphasized that these many-body effects in the polaron problem at finite temperature and finite impurity density are beyond previous single impurity theories.
While our result successfully reproduces experimental results in several regimes and predict the polaron properties which no one has reported, we found that there are still differences between theories and experiments with respect to the effective mass as well as the decay rate somehow beyond finite temperature and impurity density effects, which remain as our important future problem. 
In particular, an effect of a harmonic trap is important to compare our results with the observed rf-spectra \cite{PhysRevLett.118.083602} in detail, and our present work can include such effects by employing the local density approximation \cite{PhysRevA.96.033614}. 
It is also interesting to extend our analyses to  mass-imbalanced~\cite{Kohstall1112}
and two-dimensional systems~\cite{koschorreck2012}
already realized in ultracold Fermi gases.
\par
\ack
We thank F. Scazza, A. Recati, S. Giorgini, M. Horikoshi and Y. Nishida for useful discussions.
H. T. is supported by a Grant-in-Aid for JSPS fellows (No. 17J03975).
S. U. is supported by JSPS KAKENHI Grant Number JP17K14366.
This work was partially supported by RIKEN iTHEMS Program.
\par
\appendix
\section{Analytic continuation}
%%%%%%%%%%%%%%%%%%%%%%%%%%%%%%%%%%%%%%%%%%%%%%%%%%%%%%%%%%%%%%%%%%%%%
\begin{figure}[t]
\begin{center}
\includegraphics[width=8cm]{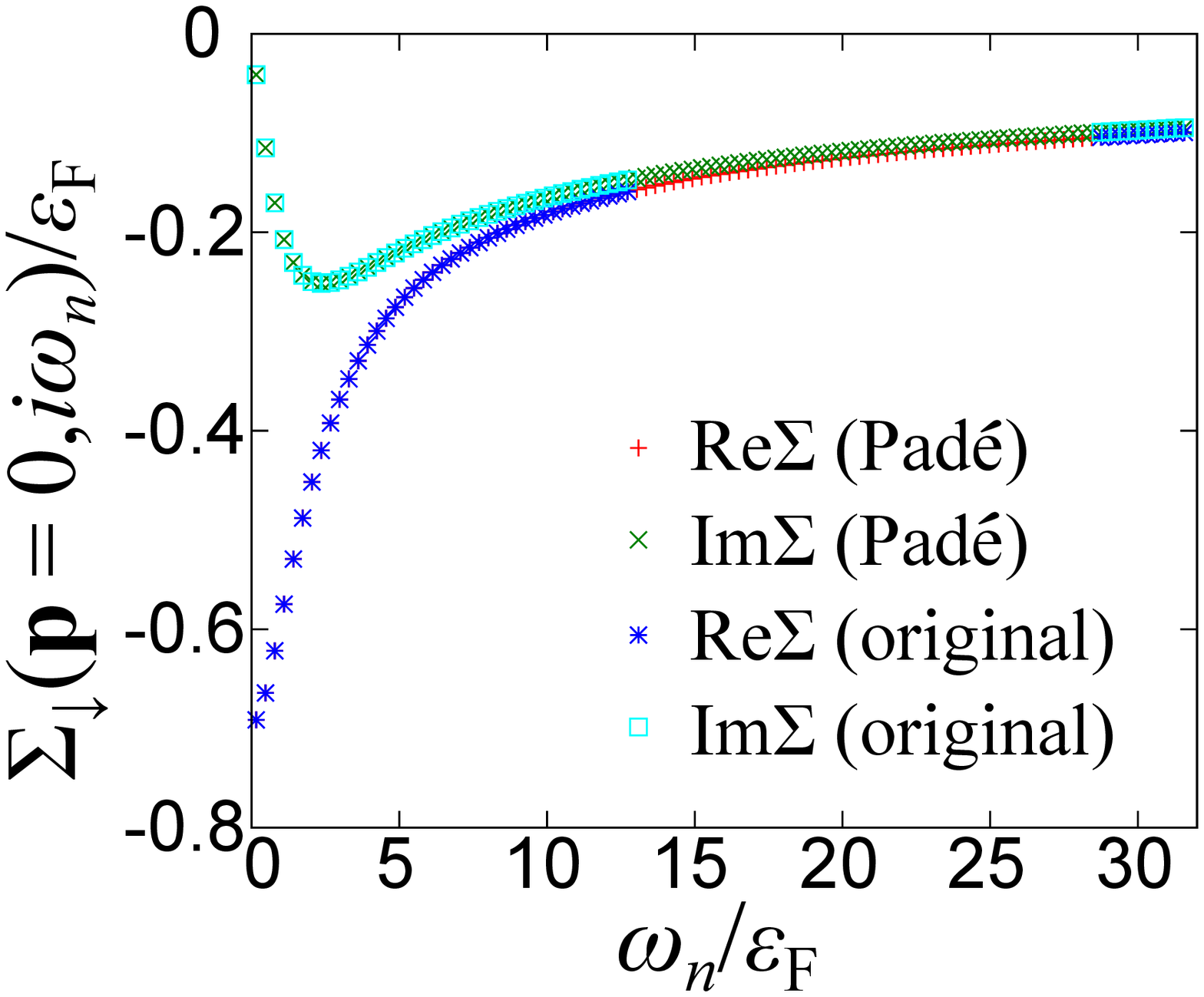}
\end{center}
\caption{Calculated impurity self-energy $\Sigma_{\downarrow}(\mathbf{p}=0,i\omega_n)$ at $T=0.05 T_{\rm F}$, $y=0.12$, and $(k_{\rm F}a_s)^{-1}=0$
and the comparison with the interpolated result obtained by the Pad\'{e} approximation
,where \textit{original} is the self-energy from Eq. (\ref{eqadd1}) without the interpolation.}
\label{figse}
\end{figure}
%%%%%%%%%%%%%%%%%%%%%%%%%%%%%%%%%%%%%%%%%%%%%%%%%%%%%%%%%%%%%%%%%%%%%
\par
In general, the analytic continuation is sensitive to noises, and theoretical approaches with statistical errors such as
Monte-Carlo methods suffers from this procedure from the imaginary time $\tau$ to the real frequency $\omega$~\cite{PhysRevA.94.051605,Goulko}. 
On the other hand, the ETMA used in this work is free from statistical errors, and therefore we can implement
the conventional numerical continuation methods.
In this work, we adopt the Pad\'{e} approximation  to examine the spectral structure in the Fermi polaron system.

In our case, the self-energy has been already calculated in the complex energy plane in terms of the Matsubara frequency located at imaginary energy axis. 
It is known that the Pad\'{e} approximation is applicable to reproduce the pole structure in this plane \cite{JLTP.29.179}.
In fact, the photoemission spectra obtained from the many-body $T$-matrix theory with the Pad\'{e} approximation
well reproduce the experimental result in a strongly interacting unpolarized Fermi gas \cite{Miki.Ota}.
In addition, the Pad\'{e} approximation has been successfully applied to the Fermi polaron system with 
the functional renormalization group~\cite{Schmidt:2011zu},
which is also free from statistical errors. 
To double-check how the Pad\'{e} approximation works well, 
we focus on the calculated self-energy in Eq. (\ref{eqadd1}).
In Fig. \ref{figse}, we show the comparison between the ETMA self-energy $\Sigma_{\downarrow}(\mathbf{p}=0,i\omega_n)$ (from $n=0$ to $n=40$ and from $n=90$ to $n=100$) and 
the interpolated results of them by means of the Pad\'{e} approximation, where the data between $n=40$ and $n=90$ are interpolated. 
One can see that the Pad\'{e} approximation smoothly interpolate the original ETMA self-energy.

\vspace{0.5cm}
\bibliographystyle{apsrev4-1}
%\bibliography{reference}

\end{document}